\documentclass[twocolumn,aps,pra]{revtex4}
\usepackage{graphicx}% Include figure files
\usepackage{amssymb}
\usepackage{amsmath}
\usepackage{float}

\newcommand{\me}{\mathrm{e}}

\begin{document}

\title{Exploiting geometric degrees of freedom in topological quantum
 computing}

\author{Haitan Xu$^1$ and Xin Wan$^{2,1}$}
\affiliation{$^1$Zhejiang Institute of
Modern Physics, Zhejiang University, Hangzhou, 310027, China}
\affiliation{$^2$Asia Pacific Center for Theoretical Physics and
Department of Physics, Pohang University of Science and
Technology, Pohang, Gyeongbuk 790-784, Korea}

\begin{abstract}
  In a topological quantum computer, braids of non-Abelian anyons in a
  (2+1)-dimensional space-time form quantum gates, whose fault
  tolerance relies on the topological, rather than geometric,
  properties of the braids. Here we propose to create and exploit
  redundant geometric degrees of freedom to improve the theoretical
  accuracy of topological single- and two-qubit quantum gates.  We
  demonstrate the power of the idea using explicit constructions in
  the Fibonacci model. We compare its efficiency with that of the
  Solovay-Kitaev algorithm and explain its connection to the leakage
  errors reduction in an earlier construction [Phys. Rev. A 78, 042325
  (2008)].

\end{abstract}

\maketitle

\section{Introduction}

Topological quantum computation is a rapidly developing subject in
recent years~\cite{kitaev03,freedman02,freedman022,dassarma07}. In
this novel scheme of quantum computation, information is stored in
topological quantum states and intrinsically protected from local
noises, and manipulation of quantum information is achieved by
topological operations. A prototypical topological quantum computer
is envisaged to be a system of exotic quasiparticles called
non-Abelian anyons,
which are believed to exist in various
two-dimensional quantum
systems~\cite{moore91,greiter92,morf98,Ed2000,wan06,wan08,feiguin,
read99,ardonne99,ardonne01,fendley05,kitaev06,sarma06}.
A multiple of these anyons with fixed coordinates span a
multi-dimensional Hilbert space, which can be used to construct
qubits or encode quantum information~\cite{nayak96,ivanov01,ardonne07}.
Schemes have been proposed to control and move anyons
microscopically~\cite{Freedman06,bonderson08,dassarma05}. The
worldlines of these anyons intertwine in (2+1)-dimensional
space-time forming braids, which are quantum gates for topological
quantum computation.

In earlier studies~\cite{bonesteel05,hormozi07,thesis,xu08},
researchers developed the method of brute-force search (and its
variant) among braids within given braid length (measured by the
number of exchanges) to achieve a generic single-qubit quantum gate in
the Fibonacci anyon model (which may be realizable in
fractional quantum Hall systems~\cite{read99,xia04}),
and then constructed controlled-rotation
gates from single-qubit gates. These works explicitly demonstrated the
equivalence between a specific theoretical realization of topological
quantum computer and a universal quantum computer
model~\cite{freedman02}. In general, a single-qubit quantum gate can
be represented by a $2\times 2$ unitary matrix
\begin{equation}
\label{eq:unitarymatrix} G=\me^{i\alpha}\left[
 \begin {array}{cc}
 \sqrt{1-b^2}\me^{-i\beta} &  b\me^{i\gamma}\\
 -b\me^{-i\gamma} & \sqrt{1-b^2}\me^{i\beta}
\end{array}\right],
 \end{equation}
where $b$, $\alpha$, $\beta$ and $\gamma$ are real parameters. Apart
from the overall phase factor $\me^{i\alpha}$, one needs three
parameters $b$, $\beta$ and $\gamma$ to specify the matrix. Within a
given braid length, there are only a finite number of topological
quantum gates, which form a discrete set in the $U(2)$ space, thus
generic gates can only be realized with a distribution (wide on
logarithmic scale) of error even in the ideal scenario (without
technical or practical hindrance), due to the discrete nature of braid
topology. This contrasts to many proposals of conventional quantum
computation, where quantum gates can be realized by continuously
tuning physical parameters so generic quantum gates are expected to be
realized with only a narrow distribution of error (due to technical
imperfections). On the other hand, the discreteness (thus error) in
the realization of quantum gates with braids of finite length shares
the same origin as the fault tolerance of topological quantum
computation, as quantum states and quantum gates (braids) are
topological and robust against local perturbations. This therefore
poses an interesting question: How can we efficiently find the braid
with finite length that approximates a desired quantum gate with error
as small as possible?

In a recent work, the authors proposed a novel construction of
low-leakage topological quantum computation based on the principle of
error reduction by error introduction~\cite{xu08}. In topological
quantum computation, the leakage errors in two-qubit gates one wants
to minimize is often of topological origin, e.g., arising from the
existence of noncomputational states. Nevertheless, one may find a
class of equivalent braid constructions characterized by an additional
geometric degree of freedom in braid segments, which can be said to
correspond to a $U(1)$ symmetry of the construction. In practice,
however, due to the discreteness of braids in the target space, such a
symmetry is merely a pseudo symmetry. One finds that some of the
constructions can have exponentially smaller errors than others --
they are exactly what we want to find. The successful application of
the principle in the Fibonacci anyon model led to the discovery of an
exchange braid (with a length of 99) that exchanges anyons between two
different qubits, which can be used to construct generic
controlled-rotation gates with leakage error as small as $10^{-9}$.
However, the idea can not be directly applied to construct
single-qubit gates because apparently there is no such geometric
freedom.

In this paper, we generalize the idea that errors in topological gate
construction can be reduced by the introduction of an additional
geometric redundancy (or symmetry)~\cite{xu08} and show that it can
also be applied to the construction of generic single-qubit gates with
unprecedented efficiency and accuracy in theory.  We demonstrate this
idea explicitly in the Fibonacci model, though it is applicable in
generic models that support universal topological quantum computation.
By introducing new degrees of freedom with unitary similarity
transformation, we show a generic single-qubit gate can be
approximated to a distance~\cite{distance} of the order $10^{-10}$ by
a braid of length $\sim$300, which is more efficient than a direct
application of the Solovay-Kitaev algorithm~\cite{hormozi07}. We also
discuss the significant reduction of leakage error by error
introduction in a parallel construction of two-qubit controlled-gates
(see also Ref.~\cite{thesis,hormozi09}) to demonstrate the power and
generality of the idea of topological error reduction by exploiting
redundant geometric degrees of freedom.

\section{Single-Qubit  Gate Construction}

Let us first discuss the high-accuracy construction of a generic
single-qubit gate $G$ represented by Eq.~(\ref{eq:unitarymatrix}).
To create additional degrees of freedom needed for error reduction,
one is tempted to decompose the target gate as $G = G_1 G_2$, where
\begin{equation}
\label{eq:unitarymatrix1} G_{1,2}=\me^{i\alpha_{1,2}}\left[
 \begin {array}{cc}
 \sqrt{1-b_{1,2}^2}~\me^{-i\beta_{1,2}} &  b_{1,2}\me^{i\gamma_{1,2}}\\
 -b_{1,2}\me^{-i\gamma_{1,2}} & \sqrt{1-b_{1,2}^2}~\me^{i\beta_{1,2}}
\end{array}\right]
\end{equation}
are unitary matrices, or quantum gates. Unfortunately, the degrees of
freedom of the two gates are dependent, as can be immediately seen
from $G_1=GG_2^+$, i.e., one of the gate (e.g., $G_1$) is completely
fixed by the other gate (e.g., $G_2$) up to an unimportant global
phase factor.  Therefore, one cannot use the new degrees of freedom
for gate optimization. In fact, this decomposition of $G$ is the
mathematical structure of the bidirectional search~\cite{xu08}, which
facilitate the search for twice longer braids without a significant
increase in CPU power.

In fact, to create separable degrees of freedom with two gates, we can
write $G = G_1G_2G_1^+$, a unitary similarity transformation,
which creates geometric redundancy. It is easier to visualize the
transformation in terms of rotation in three dimensions, thanks to the
homomorphism between the groups $SO(3)$ and $SU(2)$. This means that a
rotation around an arbitrary axis $l$ by an angle $\theta$ on a Bloch
sphere can be carried out by first rotating $l$ to another direction
$l'$, then rotating around $l'$ by an angle $\theta$, and finally
rotating $l'$ back to $l$. The geometric interpretation clearly
indicates that the freedom in the choice of $l'$ can be exploited to
optimize the single-qubit gate $G$, because these apparently
equivalent realizations use different braids to approximate the target
gate with entirely different accuracy.

We can use a phase gate $P$ (a diagonal matrix) to
illustrate the determination of $G_1$ and $G_2$ without loss of
generality. This is because, according to the spectral theorem for
normal matrices, any unitary matrix $G$ can be unitarily
diagonalized as $G = S^+ P S$, where $S$ is a unitary matrix.
We can then contract $S$ with $G_1$, so that $P = \tilde{G}_1 G_2
\tilde{G}^+_1$, where $\tilde{G}_1 = S G_1$.

For concreteness, let us assume
\begin{equation}
\label{eq:diagonal}
P=\left[
 \begin {array}{cc}
\me^{-i\beta} &  0\\
0 & \me^{i\beta}
\end{array}\right],
 \end{equation}
which is a rotation around the $z$ axis by an angle $\beta$. The
parameters $b_{1,2}$, $\beta_{1,2}$ and $\gamma_{1,2}$ of $G_1$ and
$G_2$ that decompose $P$ must, therefore, satisfy
\begin{eqnarray}
\label{eq:condition1}
&& (1-b_2^2)^{1/2}\cos\beta_2=\cos\beta, \\
\label{eq:condition2}
b_1 &=& \frac{b_2}{\sqrt{
2\sin^2 \beta +2(1-b_2^2)^{1/2}\sin\beta_2 \sin\beta } }, \\
\label{eq:condition3}
&&\beta_1+\gamma_1=\gamma_2+(k+1/2)\pi,
\end{eqnarray}
where the integer $k$ is even for positive $\sin\beta$ or odd for
negative $\sin\beta$ (we exclude the case $\sin\beta=0$ when the
corresponding gate is proportional to the identity). From
Eq.~(\ref{eq:condition1}) we can see that $G_2$ is a rotation of the
same angle as that of $P$ but around a new axis, which is related to
the original rotation axis of $P$ by Eqs.~(\ref{eq:condition2}) and
(\ref{eq:condition3}). Nevertheless, when $G_2$ is fixed, $G_1$ is
only partially determined by $G_2$ and $P$ and still has a degree of
freedom (between $\beta_1$ and $\gamma_1$).  In other words, the
similarity transformation $G_1 G_2 G_1^{+}$ has an $SU(2)$ symmetry,
from which we have three free parameters to choose. $G_1$, the
rotation of $z$-axis to a new axis, has a $U(1)$ symmetry (i.e., one
free parameter), while $G_2$, fixing the direction of the new axis,
has a symmetry of $SU(2)/U(1) \sim S^2$ (i.e., two free parameters).
Hence, we can successfully separate the three degrees of freedom
into two parts in $G_1$ and $G_2$, which allow us to efficiently
search for high-accuracy single-qubit gates.

As an explicit demonstration of the algorithm, we construct a phase
gate
\begin{equation}
\label{eq:p1} P_1=\me^{i 7 \pi/5} \left[
 \begin {array}{cc}
\me^{-i 2\pi/5} &  0\\
0 & \me^{i 2 \pi / 5}
\end{array}\right]=\left[
 \begin {array}{cc}
-1 &  0\\
0 & \me^{-i \pi / 5}
\end{array}\right],
\end{equation}
in the Fibonacci anyon model (please refer to
Refs.~\cite{xu08,Preskill} for details of this model), where there are
two types of anyons with topological charges 0 (vacuum) and 1
(Fibonacci anyon) satisfying a nontrivial fusion rule
$1\times1=0+1$. We use two pairs of Fibonacci anyons with total charge
0 to encode one bit of quantum information. The basis states are
chosen as $|0\rangle=|((11)_0(11)_0)_0\rangle$ and
$|1\rangle=|((11)_1(11)_1)_0\rangle$, where the subscripts specify the
fusion results (or total topological charges) of the anyons in the
preceding brackets. Therefore, four-strand braids can be generated by
the elementary braids with representation
\begin{eqnarray}
\label{eq:s2}
\sigma_1 = \sigma_3 &=& \left [
\begin{array}{cc}
\me^{-i 4 \pi / 5} & 0 \\
0 & -\me^{-i 2 \pi / 5}
\end{array}
\right ], \\
\sigma_2 &=& \left [
\begin{array}{cc}
-\tau \me^{-i \pi / 5} & -\sqrt{\tau} \me^{i 2 \pi / 5} \\
 -\sqrt{\tau} \me^{i 2 \pi / 5} & -\tau
\end{array}
\right ],
\end{eqnarray}
and their inverses, where $\tau = (\sqrt{5} - 1)/2$.  We find a set of
$G_1=$$\sigma_2$$\sigma_3^{-2}$$\sigma_2^{4}$$\sigma_3^{-4}$$\sigma_2^{-2}$$\sigma_3^{-4}$$\sigma_2^{2}$$\sigma_3^{-2}$$\sigma_2^{2}$$\sigma_3^{-4}$$\sigma_2^{4}$$\sigma_3^{-4}$$\sigma_2^{4}$$\sigma_3^{2}$
$\sigma_2^{2}$$\sigma_3^{-4}$$\sigma_2^{-4}$$\sigma_3^{2}$$\sigma_2^{-4}$$\sigma_3^{-2}$$\sigma_2^{2}$$\sigma_3^{4}$$\sigma_2^{-2}$$\sigma_3^{4}$$\sigma_2^{4}$$\sigma_3^{2}$$\sigma_2^{2}$$\sigma_3^{-2}$$\sigma_2^{-2}$$\sigma_3^{-2}$$\sigma_2^{4}$$\sigma_3^{-4}$
$\sigma_2^{-2}$$\sigma_3^{-2}$$\sigma_2^{-2}$$\sigma_3^{-4}$ and $G_2=\sigma_2^{4}$$\sigma_3^{4}$$\sigma_2^{-2}$$\sigma_3^{2}$$\sigma_2^{4}$$\sigma_3^{-4}$$\sigma_2^{-4}$$\sigma_3^{2}$$\sigma_2^{-4}$
$\sigma_3^{-4}$$\sigma_2^{4}$$\sigma_3^{4}$$\sigma_2^{4}$$\sigma_3^{-4}$$\sigma_2^{-4}$$\sigma_3^{4}$$\sigma_2^{4}$$\sigma_3^{4}$$\sigma_2^{2}$$\sigma_3^{-4}$$\sigma_2^{4}$$\sigma_3^{-2}$$\sigma_2^{-4}$$\sigma_3^{-4}$.
One can verify $G_1 G_2 G_1^+$, with 280
interchanges~\cite{cancellation}, approximates $P_1$ with a distance
$\sim 4 \times 10^{-10}$. In general, such a precision can be achieved
by a braid of length $\sim$300 for a generic single-qubit gate with
the algorithm specified above. A braid of similar accuracy is expected
to exist at a length of as short as 150~\cite{xu08}, but the search for
it is exponentially harder.

\section{Two-Qubit Gate Construction}

The single-qubit construction scheme echos the earlier low-leakage
two-qubit construction scheme~\cite{xu08}, in which one exchanges a
two-anyon composite in the control qubit with the neighboring anyon in
the target qubit and performs single-qubit operations on the new
target qubit, which translate into controlled rotations in the
original two-qubit system, before one exchanges the anyons back to
their original locations. There is, however, a notable difference: the
$SU(2)$ symmetry in the single-qubit construction is broken in the
two-qubit construction due to leakage error. As discussed in the
introduction, only a $U(1)$ symmetry exists in the two-qubit
construction when we enforce leakage errors to be negligible. This
suggests that {\it the requirement of zero (or ultralow) leakage
  errors in the two-qubit construction eats two degrees of
  freedom}. It becomes clear in an alternative high-accuracy
($\sim$$10^{-10}$) implementation presented in the following, in which
the construction of the two-qubit gates are based on a mapping from
two qubits to one qubit in the four-anyon encoding scheme (see
also~\cite{thesis,hormozi09}), which also demonstrates the generality
of the idea of topological error reduction by redundant geometric
degree of freedom.

\begin{figure}
\begin{center}
\includegraphics[width=8cm]{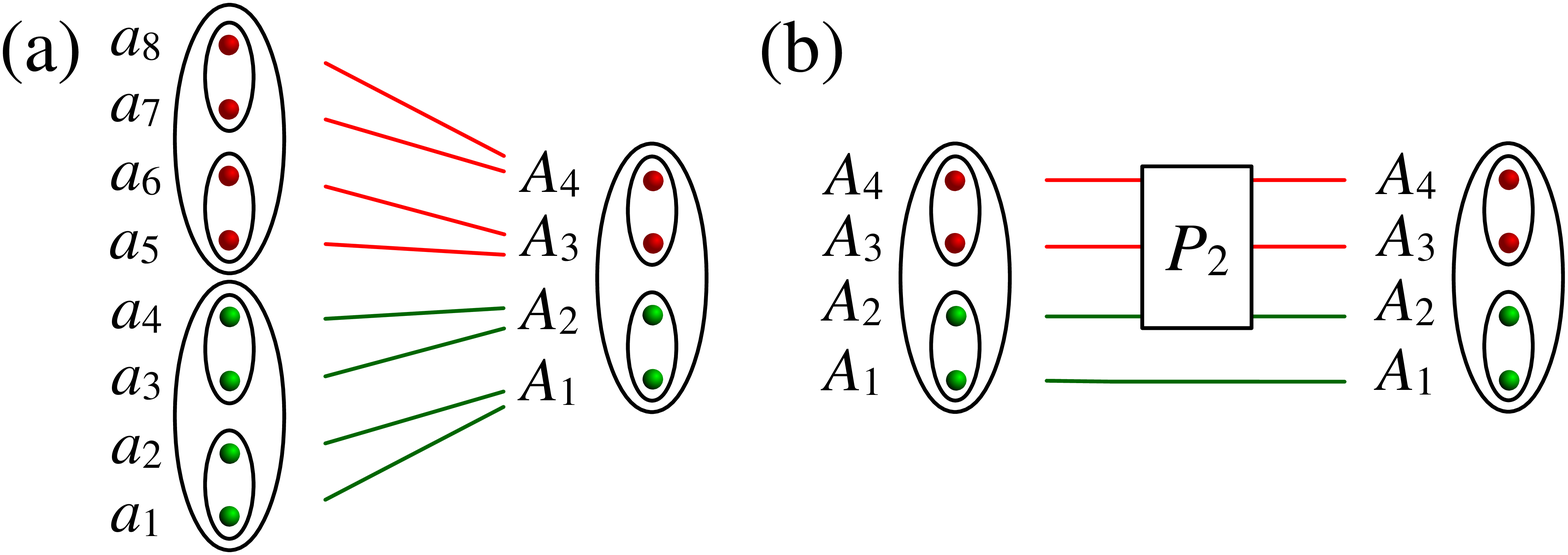}
\end{center}
\caption{ \label{fig:mapping}(a) Mapping two qubits ($a_1$-$a_4$ and
$a_5$-$a_8$) into one qubit consisting of four composite anyons $A_1$-$A_4$.
(b) Braid $P_2$ of the composite anyons $A_1$ to $A_4$. As explained in
the text, we only
move the composite anyon $A_2$ to braid with the composite anyons $A_3$
and $A_4$ and return $A_2$ back to the original position at the end of the
braid. }
\end{figure}

Explicitly, the mapping scheme is the following. For clarity, we
label the anyons in the target qubit $a_1$-$a_4$ and those in the
control qubit $a_5$-$a_8$ as in Fig.~\ref{fig:mapping}(a). We can
treat the two pairs of anyons in each qubit as two composite anyons,
which have a total topological charge 0. Then we have a mapping from
two qubits of Fibonacci anyons to one qubit of composite anyons,
which we label $A_1$-$A_4$ as in Fig.~\ref{fig:mapping}(a). The
computational basis states are chosen as
\begin{equation}
\label{eq:computationalbasis1}
\begin{split}
|00\rangle=|((11)_0(11)_0)_0 ((11)_0(11)_0)_0\rangle=|((\tilde{0}\tilde{0})_0(\tilde{0}\tilde{0})_0)_0\rangle,\\
|01\rangle=|((11)_0(11)_0)_0 ((11)_1(11)_1)_0\rangle=|((\tilde{0}\tilde{0})_0(\tilde{1}\tilde{1})_0)_0\rangle,\\
|10\rangle=|((11)_1(11)_1)_0 ((11)_0(11)_0)_0\rangle=|((\tilde{1}\tilde{1})_0(\tilde{0}\tilde{0})_0)_0\rangle,\\
|11\rangle=|((11)_1(11)_1)_0((11)_1(11)_1)_0\rangle=|((\tilde{1}\tilde{1})_0(\tilde{1}\tilde{1})_0)_0\rangle,
\end{split}
\end{equation}
where $\tilde{0}$ and $\tilde{1}$ denote the topological charge of
composite anyons. In fact, the composite qubit is not a qubit in the
normal encoding scheme, because the composite anyons $A_i$ can have
charge 0. Each pair of composite anyons (e.g., $A_1$ and $A_2$)
always have total charge 0, unless leakage error occurs so each of
the original qubits has total charge 1. Therefore, the class of braids
that we look for to manipulate the the composite
qubit as in Fig.~\ref{fig:mapping}(b) without introducing leakage
errors are the ones that realize phase gates, for example,
\begin{equation} \label{eq:P2-2}
P_2=\me^{i \alpha_2}\left[
 \begin {array}{cc}
\me^{-i\beta_2} &  0\\
0 & \me^{i\beta_2}
\end{array}\right].
\end{equation}
We see immediately that two degrees of freedom in $SU(2)$ disappear
due to the requirement of zero leakage errors and we are left with a
$U(1)$ symmetry only.  In practice, we restrict ourselves to move the
composite anyon $A_2$ to braid with the composite anyons $A_3$ and
$A_4$ and return $A_2$ back to the original position at the end of the
braid; this is known as a weave~\cite{bonesteel05}. When the composite
qubit is in the state
$|((\tilde{1}\tilde{1})_0(\tilde{1}\tilde{1})_0)_0\rangle$, this braid
will introduce a phase factor $\me^{i (\alpha_2-\beta_2)}$ to the
two-qubit system [e.g., a phase factor -1 for the braid approximating
$P_1$ in Eq.~(\ref{eq:p1})]. While if the composite qubit is
originally in the other computational states in
Eq.~(\ref{eq:computationalbasis1}), either the topological charge of
the composite anyon $A_2$ or the topological charges of the composite
anyons $A_3$ and $A_4$ are 0. Since the braid between an anyon with
topological charge 0 and another anyon with topological charge either
0 or 1 does not change the state of the system, the braid will bring
only a trivial phase factor 1 to the system. Thus a braid
approximating the single-qubit phase gate $P_2$ in Eq.~(\ref{eq:P2-2})
corresponds to a controlled-phase gate, e.g., a controlled-$Z$ gate
for the braid approximating $P_1$ in Eq.~(\ref{eq:p1}).

\begin{figure}
\begin{center}
\includegraphics[width=8.5cm]{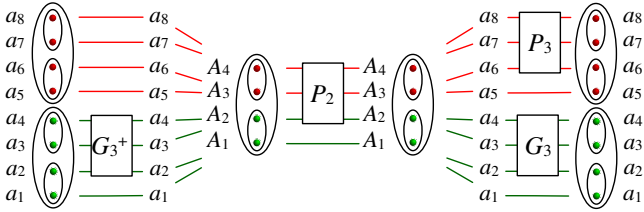}
\end{center}
\caption{\label{fig:cnot} A scheme to realize a generic
controlled-rotation gate in the Fibonacci anyon model. We consider
two qubits composed of $a_1$-$a_8$. $P_2$, which realizes a two-qubit
controlled-phase gate, is a braid acting on the
effective single qubit formed by composite anyons $A_1$-$A_4$.
$G_3$ and $P_3$ are braids of single-qubit gates that modify the
controlled-phase gate to a generic controlled-rotation gate. Note
that we choose the convention of the time direction from left to right.}
\end{figure}

A scheme to construct an arbitrary controlled-rotation gate, parallel
to the single-qubit gate construction, is illustrated in
Fig.~\ref{fig:cnot} (see also Fig.~3 in Ref.~\cite{hormozi09} for an
$SU(2)_5$ construction). In particular, we need to apply a single-qubit
gate $G_3$ on the target qubit after the controlled-phase gate and its
inverse $G_3^{-1}=G_3^{+}$ before the controlled-phase gate. This is
in the same spirit as the similarity transformation in the
single-qubit case, except that the controlled-phase gate is defined on
the composite qubit, not the target qubit. For completeness, we need
to introduce another single-qubit phase gate
\begin{equation}
\label{eq:} P_3=\me^{i \alpha_3}\left[
 \begin {array}{cc}
\me^{-i\beta_3} &  0\\
0 & \me^{i\beta_3}
\end{array}\right].
\end{equation}
to adjust the phase of the control qubit such that the resulting gate
is
\begin{equation}
\label{eq:cgateform}\me^{i
(\alpha_3-\beta_3)}\left[\begin{array}{cc|ccc}
  1&0&0&0\\
  0&1&0&0\\\hline
  0&0&&\\
  0&0&&\\
  \end{array}\right]
\put(-22,-14){\it R},
\end{equation}
where $R$ is related to $G_3$ by
\begin{equation}
\me^{i \left (\frac{\alpha_2 - \beta_2}{2} +2\beta_3 \right
)}G_3\left[
 \begin {array}{cc}
\me^{-i(\alpha_2-\beta_2)/2} &  0\\
0 & \me^{i(\alpha_2-\beta_2)/2}
\end{array}\right]G_3^+.
\end{equation}
Given a target $R$, we should search for a braid realizing the
corresponding $G_3$. As $R$ can be diagonalized by a similarity
transformation $S R S^+$, the constraint on $G_3$ is that $S G_3$
should a single-qubit phase gate with an arbitrary phase, which again
allows a redundant $U(1)$ degree of freedom.

In fact, we intentionally designed our presentation so that the braid
realization of $P_1$ in Eq.~(\ref{eq:p1}) is the same for $P_2$ in
order to construct a controlled-NOT (CNOT) gate.  Correspondingly, we
find the sequence for one instance of $G_3$ as
$\sigma_2^{-4}$$\sigma_3^{2}$$\sigma_2^{-2}$$\sigma_3^{-2}$$\sigma_2^{2}$$\sigma_3^{-2}$$\sigma_2^{-4}$$\sigma_3^{-2}$
$\sigma_2^{-4}$$\sigma_3^{-4}$$\sigma_2^{2}$$\sigma_3^{2}$$\sigma_2^{-2}$$\sigma_3^{-2}$$\sigma_2^{-2}$$\sigma_3^{-4}$$\sigma_2^{2}$$\sigma_3^{-4}$$\sigma_2^{2}$$\sigma_3^{-4}$$\sigma_2^{2}$$\sigma_3^{-2}$$\sigma_2^{-4}$$\sigma_3^{2}$$\sigma_2^{4}$
$\sigma_3^{-2}$$\sigma_2^{4}$$\sigma_3^{-4}$$\sigma_2^{4}$$\sigma_3^{2}$$\sigma_2^{-4}$$\sigma_3^{4}$$\sigma_2^{4}$$\sigma_3^{4}$$\sigma_2^{4}$.
The total braid for the CNOT gate, with an error of
$~5\times10^{-10}$, contains 280 exchanges of double braids and 208
exchanges of single braids. Note that $P_3$ is trivial for the CNOT
gate.  We would like to point out that this construction is
conceptually interesting but technically less efficient, because an
exchange of two double braids is, in fact, four exchanges of single
braids. The new single-qubit construction combined with the two-qubit
construction in the earlier proposal~\cite{xu08}, which also involves
exchanges of a single braid and a double braid, can achieve a similar
error within 1000 exchanges of single braids.

\section{Conclusion and Discussion}

In conclusion, we proposed the idea of exploiting redundant geometric
degrees of freedom in topological quantum computation to reduce the
topological errors due to discreteness of gates realized by
finite-length braids. This is possible because we can separate the
redundant degrees of freedom into (partially) independent parts, which
allows topological quantum gate construction to be more efficient.  We
also established the intriguing connection between the sacrifice of
two such degrees of freedom and the minimization of two-qubit leakage
errors.

We can understand the error reduction from a different angle. In the
three-dimensional space of unitary matrices, a target gate is just a
zero-dimensional point. The introduction of geometric redundancies
transforms the target into a one- or higher-dimensional object,
thereby allows an efficient deeper search.  The algorithm is
practically useful as computational errors can be reduced
exponentially at all length scales by the introduction of redundant
degrees of freedom as shown, e.g., in Fig.~4 of Ref.~\cite{xu08}.  The
increase in braid length by a factor of roughly three is thus well
conpensated by the exponential suppression in error.

Finding optimal braids belongs to the generic question of
approximating an arbitrary unitary operation by a set of discrete
gates (or matrices) relevant to, e.g., constructing quantum circuits
in generic quantum computation, for which a remarkable rate of
convergence can be achieved by the Solovay-Kitaev
algorithm~\cite{dawson06}.  The Solovay-Kitaev algorithm is based on
the principle of error cancellation by a group commutator structure of
$ABA^{-1}B^{-1}$ factor (given an initial $\epsilon$-net). It has been
implemented in the context of topological quantum computation by
Hormozi {\it et al.}, who achieved a gate with a distance $\simeq4.2
\times 10^{-5}$ to $iX$ with a braid of length 220 in one
iteration~\cite{hormozi07}.  Therefore, the algorithm presented here
can achieve comparable accuracy to that from applying one iteration of
the Solovay-Kitaev algorithm, albeit with braids that are about 40\%
shorter than those obtained from the Solovay-Kitaev algorithm. An
iterable modification of the algorithm, as well as its performance
comparison with the Solovay-Kitaev algorithm, is presented
elsewhere~\cite{xu09}.

\section*{Acknowledgments}

The authors thank Giuseppe Mussardo for insightful discussion.  This
work is supported by NSFC Grant No. 10504028 and the PCSIRT Project
No. IRT0754.  H.X. thanks the Asia Pacific Center for Theoretical
Physics (APCTP) for hospitality. X.W. acknowledges the Max Planck
Society and the Korea Ministry of Education, Science and Technology
for the support of the Independent Junior Research Group at
APCTP. X.W. thanks SISSA for hospitality during the write-up of this
paper.


\begin{thebibliography}{99}
\bibitem{distance}
The distance (sometimes referred as error) between two gates (or
their matrix representations) $U$ and $V$ is the operator norm
distance $d\left( U,V\right) \equiv \Vert U - V \Vert = \sup _{\Vert
\psi \Vert = 1} \Vert \left( U-V\right) \psi \Vert$.
\bibitem{cancellation}
There is a cancellation of 12 anyon interchanges when $G_2$
  and $G_1^{+}$ join.
\end{thebibliography}
\end{document}